\documentclass[3p, twocolumn]{elsarticle}
\usepackage[normalem]{ulem}  
\usepackage{color} 
\usepackage[hang,small,bf]{caption}
\usepackage[subrefformat=parens]{subcaption}
\usepackage{bm}

\renewcommand\sout{\bgroup \color{red} \ULdepth=-.5ex \ULset}

\journal{Physics Letters B}

\usepackage{graphicx}
\usepackage{amssymb}
\usepackage{amsmath}
\usepackage{color}

\newcommand{\ba}{\[\begin{aligned}}
\newcommand{\ea}{\end{aligned}\]}
\newcommand{\al}[1]{\begin{align}#1\end{align}}

\begin{document}

\begin{frontmatter}

\title{$d^\ast (2380)$ dibaryon from lattice QCD}

\author[a,b]{Shinya~Gongyo}
\ead{shinya.gongyo@riken.jp}
\author[c,a]{Kenji~Sasaki}
\author[a]{Takaya~Miyamoto}
\author[c,d,a]{Sinya~Aoki}
\author[a,b]{Takumi~Doi}
\author[b]{Tetsuo~Hatsuda}
\author[e,a]{Yoichi~Ikeda}
\author[f,a]{Takashi~Inoue}
\author[g,a]{Noriyoshi~Ishii}
\author{\\(HAL QCD Collaboration)}
\address[a]{Quantum Hadron Physics Laboratory, RIKEN, Saitama 351-0198, Japan}
\address[b]{iTHEMS Program, RIKEN, Saitama 351-0198, Japan}
\address[c]{Center for Gravitational Physics, Yukawa Institute for Theoretical Physics, Kyoto University, Kyoto 606-8502, Japan}
\address[d]{Center for Computational Sciences, University of Tsukuba, Ibaraki 305-8571, Japan}
\address[e]{Department of Physics, Kyushu University, Fukuoka 819-0395, Japan}
\address[f]{College of Bioresource Science, Nihon University, Kanagawa 252-0880, Japan}
\address[g]{Research Center for Nuclear Physics (RCNP), Osaka University, Osaka 567-0047, Japan}

\begin{abstract}
  The $\Delta\Delta$ dibaryon resonance  $d^\ast (2380)$ with $(J^P, I)=(3^+, 0)$ 
   is studied theoretically on the basis of  the 3-flavor lattice QCD simulation with heavy pion masses ($m_\pi =679, 841$ and $1018$ MeV).
  By using the HAL QCD method, the central $\Delta$-$\Delta$ potential in the ${}^7S_3$ channel is 
   obtained from the lattice data with the lattice spacing $a\simeq 0.121$ fm and the lattice size $L\simeq 3.87$ fm.
 The resultant potential shows a strong short-range attraction, so that a quasi-bound state corresponding to $d^\ast (2380)$ is formed
 with the binding energy $25$-$40$ MeV below the $\Delta\Delta$ threshold for the heavy pion masses.
 The tensor part of the transition potential from $\Delta\Delta$ to $NN$ is also  extracted to investigate the coupling strength between
 the $S$-wave $\Delta\Delta$ system with $J^P=3^+$ and the  $D$-wave $NN$ system.
 Although the  transition potential is strong at short distances, the decay width of $d^\ast (2380)$  to  $NN$ in the $D$-wave 
 is kinematically suppressed, which justifies our single-channel analysis at the range of the pion mass explored in this study.

 \end{abstract}
\end{frontmatter}

\newpage
\section{Introduction}
\label{sec:introduction}
Recently much interest has been attracted to decuplet-decuplet dibaryons  as well as to octet-octet and octet-decuplet dibaryons  
 \cite{Gal:2015rev, Clement:2016vnl, Gongyo:2017fjb,Iritani:2018sra,Sasaki:2019qnh, Morita:2019rph, Tolos:2020aln, Aoki:2020bew}.  
Theoretically, the quark Pauli principle provides  an important guideline to identify possible dibaryon channels \cite{Oka:2000wj,Inoue:2011ai}:  
If  the overlap of the quark wave functions is forbidden by the  quark Pauli principle, it 
  is  difficult  to form  dibaryons, while if the overlap is allowed or only partially forbidden, there is a chance.

To see the role of quark Pauli principle  more explicitly  in the decuplet-decuplet system, let us consider its irreducible representation of the SU(3) flavor symmetry,
\al{{\rm   {\bf 10} \otimes {\bf 10} = ({\bf 28} \oplus {\bf 27})_{{sym.}} \oplus ({\bf 35} \oplus {\bf 10}^* )_{{anti\mathchar`-sym.}}  } , \notag}
where ``sym." and ``anti-sym." stand for the flavor symmetry under the exchange of two baryons.
 Then one finds that there are two Pauli-allowed $S$-wave states:
Spin 0 in symmetric ${\bf 28}$ representation and spin 3 in anti-symmetric ${\bf 10}^*$ representation. The $\Omega\Omega$ system in the spin-0 channel  
belongs to the former, while the $\Delta\Delta$ system in the spin-3 and isospin-0 channel belongs to the latter \cite{Dyson:1964xwa}.
 In fact, these two systems have been studied extensively  by using phenomenological models (see e.g.
 \cite{Zhang:1997ny,Zhang:2000sv,Wang:1992wi, Wang:1995bg} for  the $\Omega\Omega$, and \cite{Kamae:1976at,Oka:1980ax, Yuan:1999pg, Dong:2018ryf} for the $\Delta\Delta$). 
 Only recently,  the first principle lattice QCD simulation of the baryon-baryon interactions near the physical point became possible thanks to the
   HAL QCD method, and it was shown that   the $\Omega\Omega$ interaction in the spin-0 channel supports      a shallow dibaryon state, the di-Omega,  near unitarity \cite{Gongyo:2017fjb}. 
      It is also proposed to search  for such a state by   the momentum correlation of $\Omega$-pairs   in future heavy-ion collision experiments
       \cite{Morita:2019rph}.

As for the $\Delta\Delta$ system, a dibaryon with spin-3 and isospin-0 has been  reported experimentally  \cite{Kamae:1976as,Adlarson:2011bh}.
 It is now called $d^\ast (2380)$ and has a resonance peak about 80 MeV below the $\Delta\Delta$ threshold with the total width 
 $\Gamma \simeq 70 \mathrm{MeV}$. The recent exclusive experiment has revealed its detailed properties such as the branching ratios
   into $NN$, $NN\pi$, and $NN\pi\pi$ \cite{Clement:2016vnl, Bashkanov:2015xsa}. Thus it is highly desirable to make a first principle
   lattice QCD calculation of  $d^\ast (2380)$.  However, it is a much involved task in comparison to di-Omega primarily because
    $d^\ast (2380)$ is a resonance above multi-particle thresholds such as $NN\pi$ and $NN\pi\pi$. 
     Instead of studying the problem with extensive coupled-channel approach on the lattice,  
    we take  heavy quark masses to capture the essential mechanism of the formation of $d^\ast (2380)$ from two $\Delta$s. 
    In such a lattice setup,  $\Delta$ becomes a stable particle without decaying into $N\pi$ and 
      $d^\ast (2380)$ may appear as a  spin-3 and $S$-wave quasi-bound state of $\Delta\Delta$ which can decay  to $NN$ only  through the $D$-wave and the $G$-wave.
   Such a lattice result  not only reveals  the physics behind $d^\ast (2380)$  but also  provides useful 
   input to   the effective field theory approach toward the physical point \cite{Haidenbauer:2017sws}.

This paper is organized as follows. In Sec.\ref{Sec:HALmethod}, we introduce the HAL QCD method to extract the $\Delta$-$\Delta$ central potential from lattice QCD. In Sec. \ref{Sec:Setup}, we summarize setup of our lattice QCD simulations. In Sec. \ref{Sec:CentralPot}, we show the numerical results of $\Delta$-$\Delta$ central potential in ${}^7S_3$ channel.  
 Sec. \ref{Sec:Summary} is devoted to summary. In \ref{Sec:TransitionPot}, we
 show the transition potential from $\Delta\Delta$ to $NN$ and estimate the decay rate to be small, which justifies the single-channel
 approach.

\section{HAL QCD method for  $\Delta$$\Delta$ interaction}
\label{Sec:HALmethod}
In QCD, the $\Delta$$\Delta$ potential in the $^7S_3$ channel is obtained from the equal-time Nambu-Bethe-Salpeter (NBS) wave function defined by
\al{
\psi_{n}^{\Delta\Delta} (\vec{r}) = \left< 0 \right| \left[\Delta\Delta\right]^{(s=3, I=0)}(\vec{r},0)\left| W_n; J=3, I=0 \right>, 
}
where $\left| W_n; J=3, I=0 \right>$ stands for a QCD eigenstate which has  the total energy $W_n = 2\sqrt{k_n^2+ m_\Delta^2}$ with $m_\Delta$ being $\Delta$-baryon's mass, the total spin $J=3$ and  the isospin $I=0$. $\left[\Delta\Delta\right]^{(s=3,I=0)}(\vec{r},t) = \sum_{\alpha,\beta,l,m,A,B, \vec{x}}P^{(s=3, I=0)}_{\alpha,\beta, l, m, A,B}\Delta_{\alpha,l}^A (\vec{x}+ \vec{r}, t)\Delta^B_{\beta,m}(\vec{x},t)$ is a two $\Delta$-baryon operator with $P^{(s=3, I=0)}_{\alpha,\beta, l, m, A,B}$ being the projection operator onto the internal spin $s=3$ and $I=0$. The $\Delta$-baryon operator
$\Delta_{\alpha,l}^A (\vec{x}+ \vec{r})$ with the charge index $A$, the spinor index $\alpha$, and  
the Lorentz index $l$ is constructed from the liner combinations of interpolating operators, 
$\epsilon^{abc}q_{a}^T(x)C\gamma_l q_b(x)q_{c\alpha}(x)$
with $q=u,d$ and $C\equiv \gamma_4 \gamma_2$.

We first assume that the couplings of $\Delta\Delta ({}^7S_3)$ to the $D$-wave and the $G$-wave $NN$ states below the $\Delta\Delta$ threshold 
is small and consider the single channel analysis between $\Delta$s.  Justification of this assumption will be 
discussed in \ref{Sec:TransitionPot}.  

 Since  the NBS wave function in the asymptotically large distance 
 is identical to  that of the scattering state or bound state in 2-body quantum mechanics \cite{Aoki:2013cra,Gongyo:2018gou}, 
  one can define  the $\Delta\Delta$ potential  via the Schr\"{o}dinger-type equation obtained 
   from the equal-time  NBS equation as \cite{Aoki:2009ji}:
\al{
&-\frac{\nabla^2}{m_\Delta}  \psi_{n}^{\Delta\Delta}  (\vec{r}) + \int U^{\Delta\Delta}(\vec{r},\vec{r'})\psi_{n}^{\Delta\Delta}  (\vec{r'})d^3\vec{r'}  \notag \\
&= E_n \psi_{n}^{\Delta\Delta}  (\vec{r}), \label{eq:Schredinger}
}
with $m_\Delta$ being the mass of $\Delta$
 and $E_n= k_n^2/m_\Delta$. Note that the non-local potential $U^{\Delta\Delta}(\vec{r},\vec{r'})$ is energy-independent.
The NBS wave function is related to the reduced four-point function, 
\al{
R^{\Delta\Delta}_{J=3}(\vec{r},t) &= \left<0\right| \left[\Delta \Delta\right]^{(s=3, I=0)}(\vec{r},t) \bar{J}^{J=3}_{\Delta\Delta}(0)\left|0\right>/e^{-2m_\Delta t} \notag \\
&= \sum_n a_n \psi^{\Delta\Delta}_{n}(\vec{r})e^{-\delta W_n t} + O(e^{-\Delta E^\ast \cdot t})
}
with $a_n = \left<W_n;J=3,I=0\right| \bar{J}^{J=3}_{\Delta\Delta}(0)\left| 0\right>$, $\delta W_n = W_n - 2m_\Delta$, $\Delta E ^\ast (>0)$ being the energy difference between the inelastic threshold and  $2m_\Delta$, and $\bar{J}^{J=3}_{\Delta\Delta}(0)$ being a source operator with $J=3$. 

Below the inelastic threshold, $R^{\Delta\Delta}_{J=3}(\vec{r},t)$ satisfies 
 the  time-dependent HAL QCD equation  \cite{HALQCD:2012aa},
 \al{
\left(\frac{\nabla^2}{m_\Delta}-\frac{\partial}{\partial t} + \frac{1}{4m_\Delta}\frac{\partial ^2}{\partial t^2}\right)R^{\Delta\Delta}_{J=3}(\vec{r},t) \notag \\
= \int U^{\Delta\Delta}(\vec{r},\vec{r'})R_{J=3}^{\Delta\Delta}(\vec{r'},t)  d\vec{r'}.}
Using the derivative expansion of the non-local potential, $U^{\Delta\Delta}(\vec{r},\vec{r'})= V^{\Delta\Delta}(\vec{r})\delta(\vec{r}-\vec{r'})+ O(\vec{\nabla})$, the leading-order (LO) local potential can be obtained as
\al{
&V^{\Delta\Delta}(\vec{r}) \notag  \\
&=  \left[R_{J=3}^{\Delta\Delta}(\vec{r}, t)\right]^{-1}  \left(\frac{\nabla^{2}}{m_{_\Delta}}-\frac{\partial}{\partial t}+\frac{1}{4m_{_\Delta}}\frac{\partial^{2}}{\partial t^{2}}\right)R_{J=3}^{\Delta\Delta}(\vec{r}, t). \label{eq:Potential}
}

The resultant potential can then be used to  calculate the observables such as the 
 binding energy and the phase shift in the infinite volume.
 \footnote{We have observed that if one applies L\"uscher's finite volume analysis \cite{Luscher:1990ux} without a variational method, the plateaux of the two-baryon spectrum are achieved at physically unrealizable time and therefore are plagued by unresolved systematic uncertainties \cite{Iritani:2016jie, Iritani:2017rlk,Iritani:2018vfn}.  Recent results on the two-nucleon system by using  the  L\"uscher's finite volume analysis with the variational method   \cite{Horz:2020zvv}  support this view independently.  Therefore, we only report results with the HAL QCD method in this paper.}

The systematic error in Eq.(\ref{eq:Potential}) originating from the LO truncation of the derivative expansion 
 can be estimated from the residual  time-dependence of $V^{\Delta\Delta}(\vec{r})$. 
Also, the higher-order terms can be determined by using the  multiple source functions for  $\bar{J}^{J=3}_{\Delta\Delta}$.
  It was shown in  \cite{Iritani:2018vfn,Iritani:2018zbt} that the next-to-LO potential obtained by combining a wall source and a smeared source for a two-octet baryon system gives negligible effects to physical observable at low energies for heavy pion masses.
\section{Simulation setup}
\label{Sec:Setup}
We employ the full QCD gauge configurations in the flavor-$SU(3)$ limit with the renormalization-group improved gauge action and the non-perturbatively $O(a)$ improved Wilson quark action at $\beta= 1.83$ and $\kappa_{uds} =0.13710,0.13760,0.13800$ for $32^3\times 32$ lattice. The lattice spacing $a$ and  the physical volume
correspond to 0.121fm and $(3.87 \mathrm{fm})^3$, respectively. We have used 360 configurations for  $\kappa_{uds} =0.13710,0.13800$ and 480 configurations  for $\kappa_{uds} =0.13760$ given in Ref. \cite{Inoue:2011ai}.  The wall-type quark source with the Coulomb gauge fixing is employed.

To increase the statistics, the forward and backward propagations are averaged and the rotational symmetry on the lattice (4 rotations) and the translational invariance for the source position (32 temporal positions) are utilized for each configuration. The hadron masses obtained by the single exponential fit are summarized in Table I. 
The statistical errors are estimated by  the Jackknife method with 18 samples for $\kappa_{uds} =0.13710, 0.13800$ and 24 samples for $\kappa_{uds} =0.13760$. 
The fit results are slightly different from Ref. \cite{Inoue:2011ai}, because we use more statistics and different fit ranges.  
 In all cases,  $m_\Delta$ is below the threshold, $m_\pi + m_N$, so that $\Delta$ is a stable baryon.

\begin{table}[ht]
\caption{The hadron masses obtained from the single exponential fit
in the intervals,  $t/a = 6-11$ (pion) and $t/a= 7-12$ (baryons).}
\begin{center}
\begin{tabular}{cccc}
\hline \hline
$\kappa_{uds} $ &
  $m_{\pi} [{\rm MeV}]$ & $m_{N} [{\rm MeV}]$ & $m_{\Delta} [{\rm MeV}]$  \\
\hline
    \ 0.13710 & \ 1017.5(2) & \ 2019.4(5)   &  2213.6(7) \\
  \ 0.13760 & \ 840.6(2) & \ 1739.1(5)   &  1940.3(6) \\
   \ 0.13800 & \ 679.0(2) & \ 1476.9(5)   &  1676.9(8) \\
\hline \hline
\end{tabular}
\end{center}
\label{TAB.HADRON_MASS}
\end{table}

\section{$\Delta$$\Delta$ potential, phase shift, and binding energy}
\label{Sec:CentralPot}

\begin{figure}[t]
\includegraphics[keepaspectratio, scale=0.22]{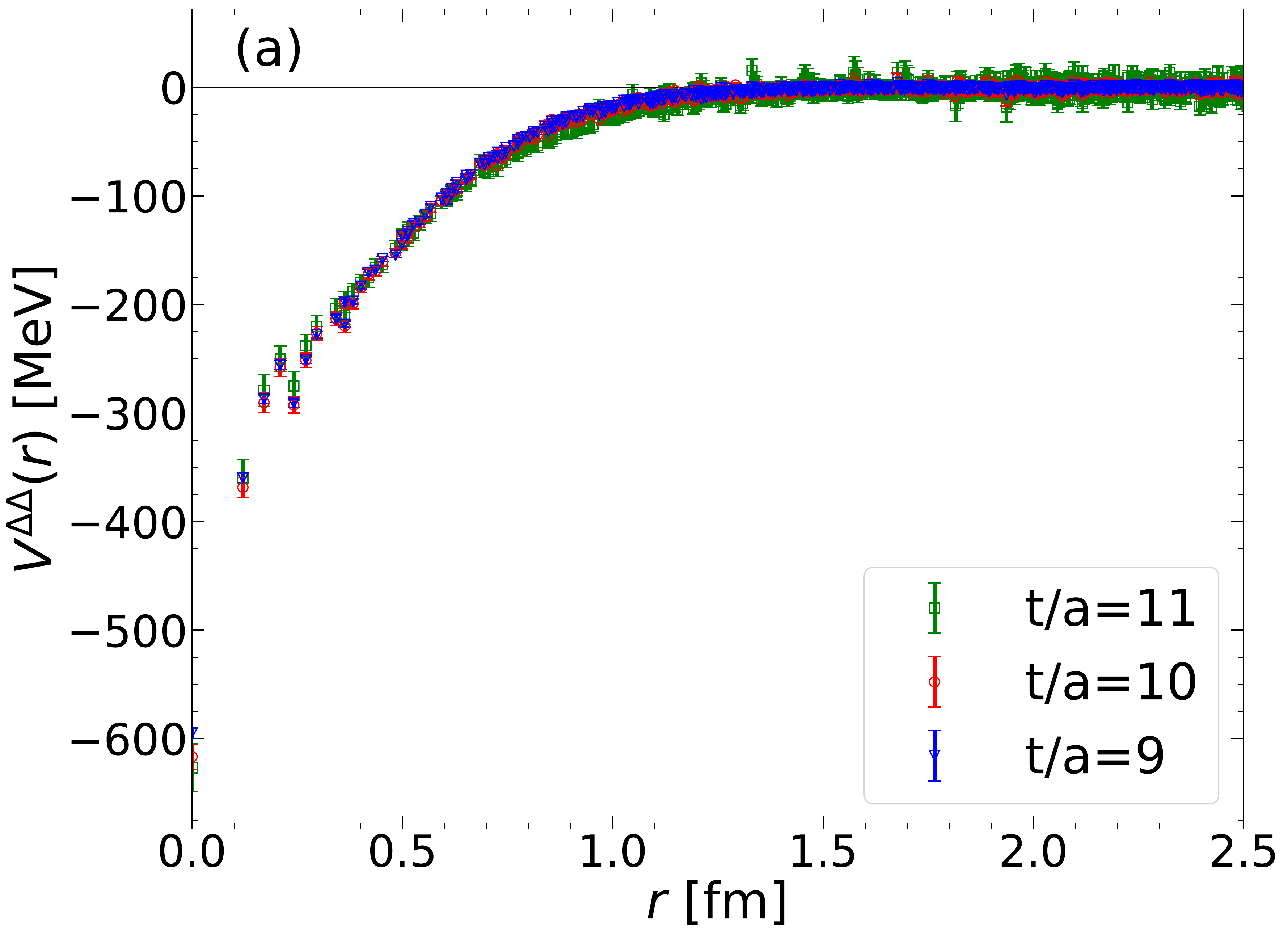} \\
\includegraphics[keepaspectratio, scale=0.22]{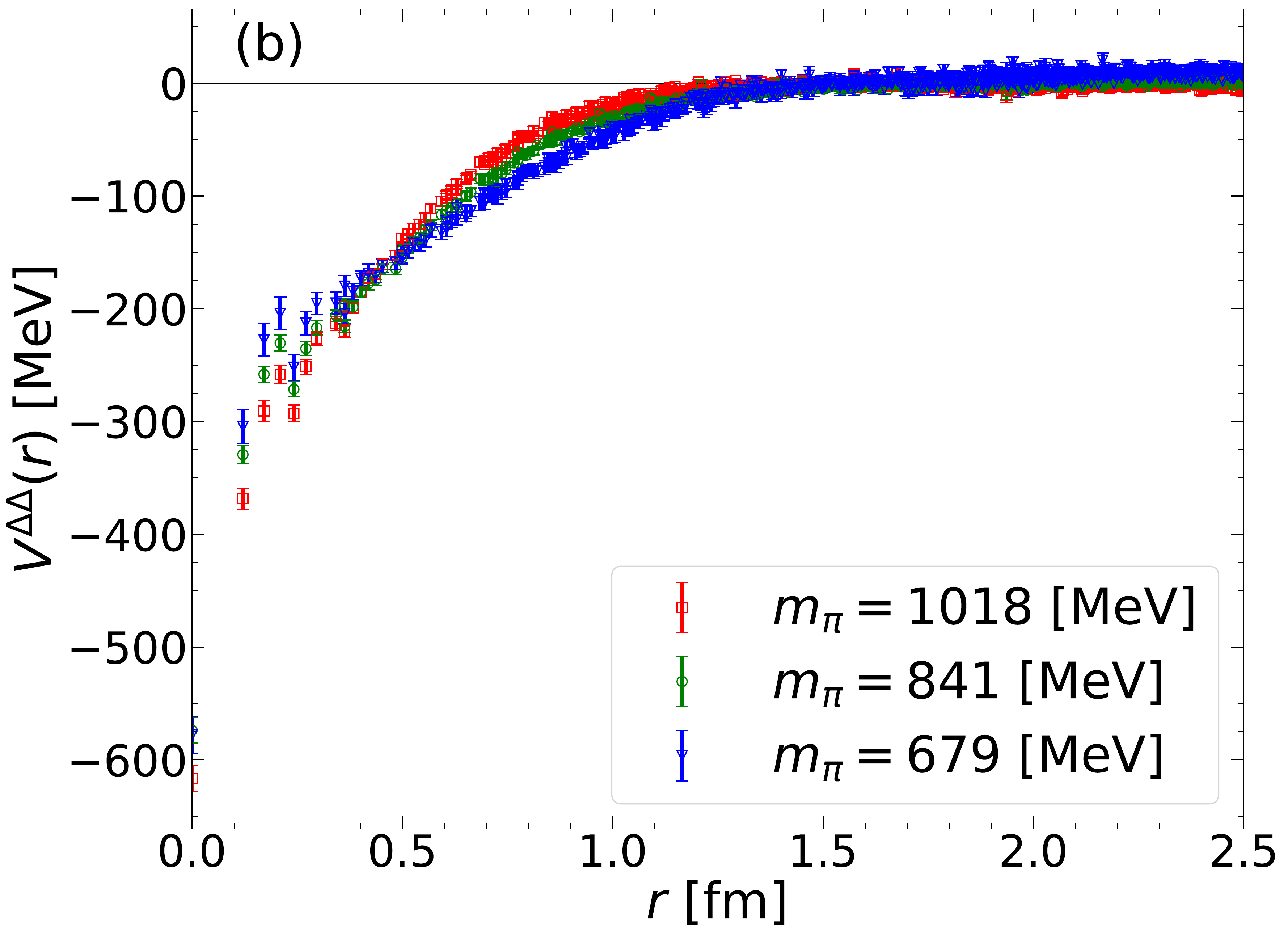}
\caption{The $\Delta$$\Delta$ central potential $V^{\Delta\Delta}(r)$ in the $^7S_3$ channel. 
(a) Results at $t/a= 9, 10, 11$ and $m_\pi = 1018 \mathrm{MeV}$.
(b) Results at $m_{\pi}=1018 \mathrm{MeV},  841 \mathrm{MeV}, 679 \mathrm{MeV}$
and $t/a = 10$.}
\label{fig:Potential}
\end{figure}

Shown in Fig.\ref{fig:Potential} are 
 the central potentials  in the $^7S_3$ channel  $V^{\Delta\Delta}(r)$  as  a function of $r$
  in the range  $t/a=9, 10, 11$ and $m_{\pi} = 679, 841, 1018$ MeV. 
  As seen from Fig.\ref{fig:Potential} (a),  $V^{\Delta\Delta}(r)$  for different $t$ are nearly identical within the 
   statistical errors indicating  that the contribution from higher-order  potential is not relevant.    
   We also find that $V^{\Delta\Delta}(r)$  is attractive for the whole distance.\footnote{
   None-smooth behavior of the potential at $r < 0.2\ {\rm fm}$ originates most likely 
   from the  lattice discretization: To remove the error, we need to take  the continuum limit by collecting
    the data for different lattice spacings.}
   The long-range part of the attraction becomes stronger as $m_{\pi}$ decreases as seen from  Fig.\ref{fig:Potential}(b).
    These features can be understood by (i)  the absence of Pauli exclusion effect for quarks in this channel, 
 (ii)  the absence of the color magnetic effect in one-gluon exchange at short distance  \cite{Oka:2000wj}, and (iii) the attractive  one-pion exchange at long distance.
We perform uncorrelated fit for the lattice data of the $\Delta\Delta$ potential  
 in the range $r=0-1.5$ fm by two Gaussians plus one Yukawa form with a form factor as 
 \al{
&V^{\Delta\Delta}(\vec{r}) \notag  \\
&= b_1 e^{- (\frac{r}{b_2})^2}+ b_3 e^{- (\frac{r}{b_4})^2}+b_5(1-e^{-(\frac{r}{b_6})^2})\frac{e^{-m_\pi r}}{r}. \label{eq:Fit}
}
For example, the fitting at $m_\pi = 679 \mathrm{MeV}$ and $t/a=10$ results in $b_1= -457 (29) \mathrm{MeV}$, $b_2= 0.090(4)\mathrm{fm}$, $b_3=-121(31) \mathrm{MeV}$, $b_4=0.15(2) \mathrm{fm}$, $b_5= -1924(533)\mathrm{MeV\cdot fm}$, $b_6= 0.98(16) \mathrm{fm}$,
with $\chi ^2/\mathrm{dof} \simeq 1$. The systematic errors in the fitting form are negligible in comparison with statistical errors and systematic errors from the $t$ dependence.

\begin{figure}[t]
\begin{center}
\includegraphics[scale=0.22]{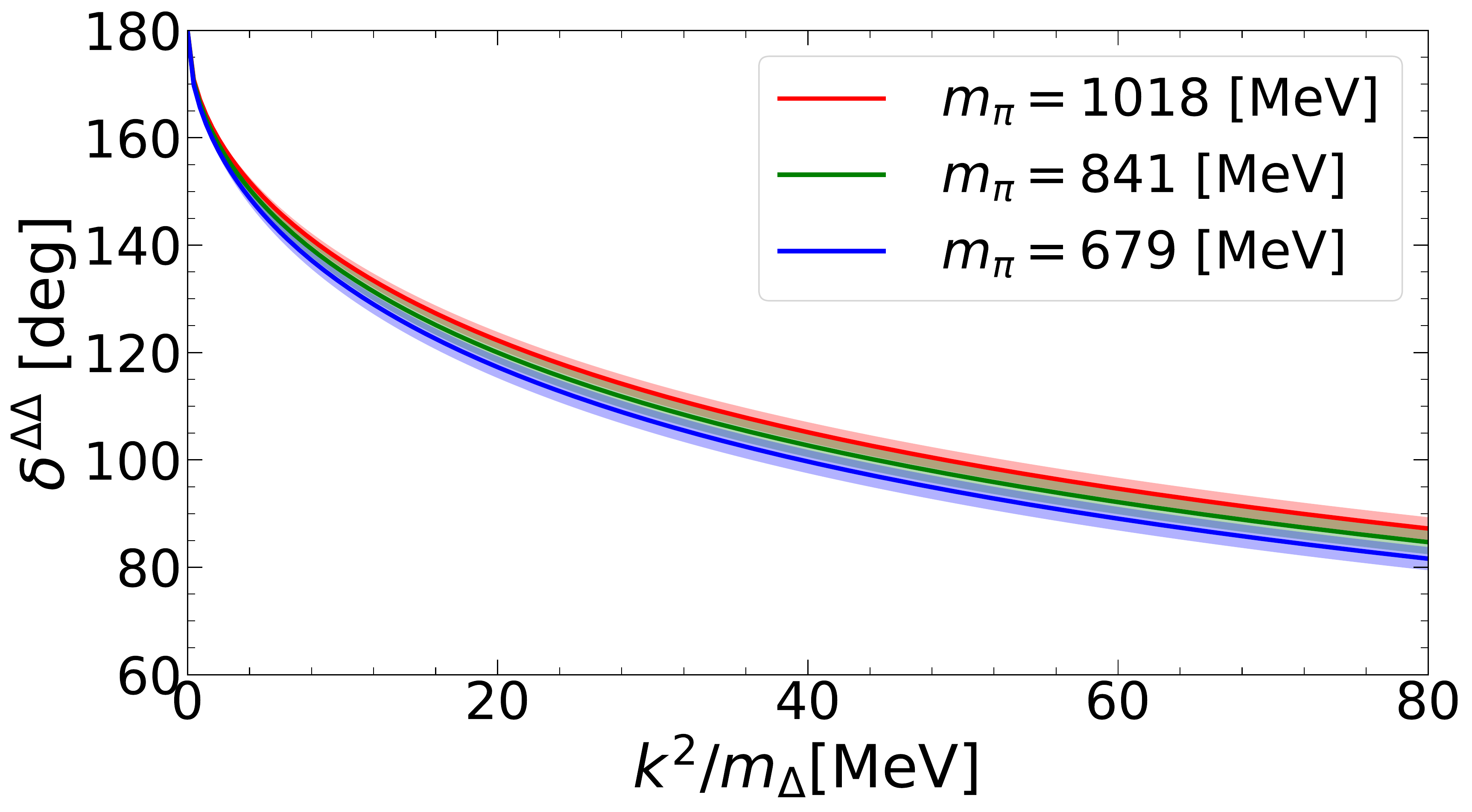}
\caption{The phase shift $\delta^{\Delta\Delta}$   in the $\Delta\Delta (^7S_3)$  channel
as a  function  of $k^2/m_{_\Delta}$ for three  pion masses.}
\label{fig:Phase-shift}
\end{center}
\end{figure}

Using  the fitted potential and solving the Schr\"odinger equation in the infinite volume,
we obtain the hypothetical $\Delta\Delta$ scattering phase shift $\delta^{\Delta\Delta}$ in the $^7S_3$ channel 
as a function of  $k^2/m_{_\Delta}$ in Fig.\ref{fig:Phase-shift} 
for three different pion masses.
  In all three cases, the phase shift starts from $180 ^\circ$ at $k^2 =0$, indicating the presence of a 
  quasi-bound state in the $\Delta\Delta (^7S_3)$  channel.

\begin{figure}[t]
\begin{center}
\includegraphics[keepaspectratio, scale=0.22]{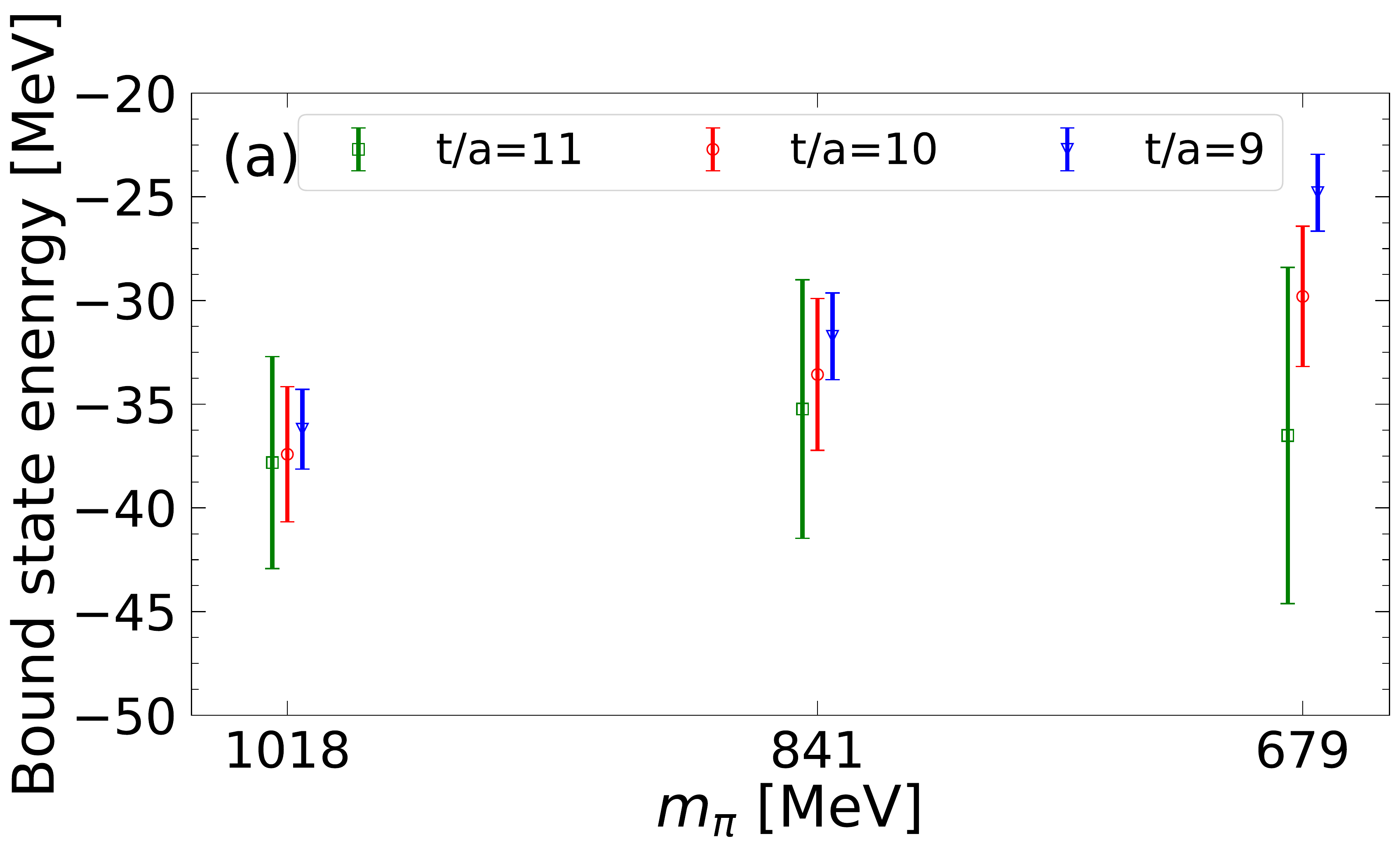} \\
\includegraphics[keepaspectratio, scale=0.22]{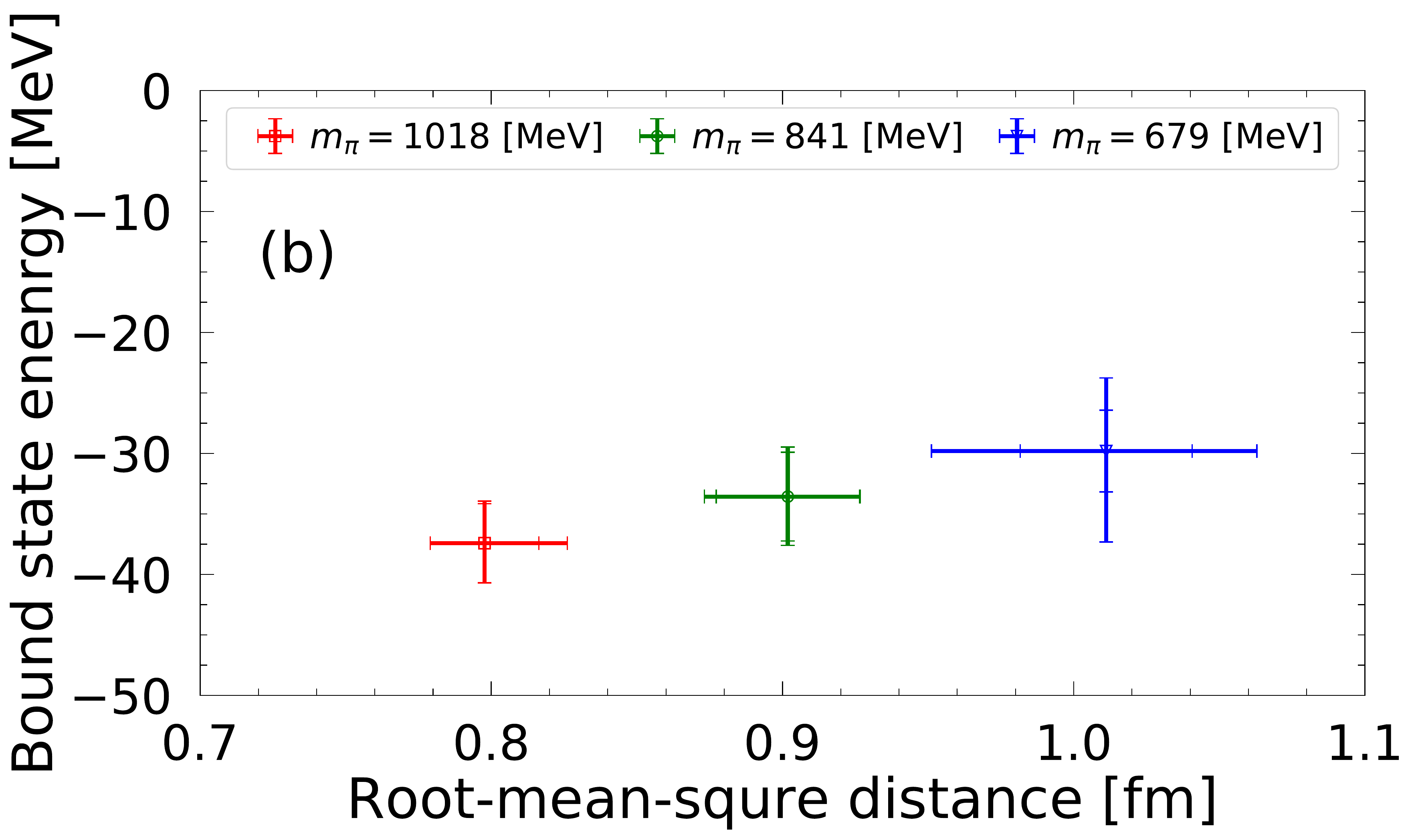} \\
\caption{(a) Bound state energy in the  $\Delta\Delta({}^7S_3)$  channel at $t/a=9,10, 11$ and  $m_{\pi}=1018 \mathrm{MeV},  841 \mathrm{MeV}, 679 \mathrm{MeV}$. (b) Bound state energy and the root-mean-square distance at $t/a=10$ and  $m_{\pi}=1018 \mathrm{MeV},  841 \mathrm{MeV}, 679 \mathrm{MeV}$. 
 Inner bars correspond to the statistical errors, while the outer bars are obtained by the quadrature of the statistical and systematic errors estimated from the central values for $t/a = 9, 11$.}
\label{fig:Binding}
\end{center}
\end{figure}

The binding energy  $B_{\Delta\Delta}$ can be also obtained from the Schr\"odinger equation. 
 The results of the bound state energy $E_0=- B_{\Delta\Delta}$  for different $t/a$ and $m_{\pi}$ are shown in Fig.\ref{fig:Binding}(a).
 Also shown in  Fig.\ref{fig:Binding}(b) are the bound state energy $E_0$  and 
  the root-mean-square distance $\sqrt{\left< r^2\right>_{\Delta\Delta}}$ of the $\Delta\Delta$ quasi-bound state.
   The typical size of the quasi-bound state is $0.8-1$ fm and the final values of the  binding energies read
\al{
m_{\pi} = 1018 \ {\rm MeV}:  & \ \ B_{\Delta\Delta} =  37.4 (3.3)(^{+1.2}_{-0.4})\  {\rm MeV}, \notag  \\
m_{\pi} = \ 841\ {\rm MeV}:   & \ \  B_{\Delta\Delta} = 33.6 (3.7)(^{+1.8}_{-1.7})\  {\rm MeV},   \notag \\
m_{\pi} = \ 679\ {\rm MeV}:  & \ \  B_{\Delta\Delta} =  29.8 (3.4)(^{+6.7}_{-5.0})\   {\rm MeV},   
\label{eq:Binding}
} with the statistical errors (first) and systematic errors from the $t$ dependence (second). 

\section{Summary}
\label{Sec:Summary}
We have studied the $\Delta\Delta$ system in the $(J,I)=(3,0)$ channel, where the resonant dibaryon $d^\ast (2380)$ was observed,
 from the lattice QCD simulation with heavy quark masses in the flavor-$SU(3)$ limit. The $\Delta$-$\Delta$ central potential in the ${}^7S_3$
 channel calculated by the HAL QCD method is found to be attractive in all distance. The phase shifts obtained by solving the Schr\"odinger 
 equation using the potential show the presence of the deep quasi-bound state below the $\Delta\Delta$ threshold. The energy below the threshold is estimated from $t/a=10$ to be about $30 \mathrm{MeV}$ 
 in the case of the lightest pion mass $m_\pi = 679 \mathrm{MeV}$.
 
 Our result implies that other members of ${\bf 10}^*$ representation such as $\Delta \Sigma ^\ast$ in the $(J,I)=(3,1/2)$ channel and $\Delta \Xi ^\ast$ in the $(J,I)=(3,1)$ channel may have  dibaryons due to the similar central attraction shown in the $\Delta$$\Delta$ system. However, the systems are more intricate even with heavy quark masses, because of their decay not only into 
 octet-octet systems but also into  octet-decuplet systems  through $D$ and $G$ waves.

 The lattice simulation of the $\Delta\Delta$ system near the physical point is left for future studies.
 Since $\Delta$ baryon can decay into $N\pi$, the $\Delta\Delta$ system can also decay into $NN\pi$ and $NN\pi\pi$
 as well as $NN$. Therefore, the coupled channel equations associated with three and four hadron systems,
 which is challenging not only in the simulation but also in the formulation on the lattice \cite{Aoki:2012bb}, are needed to extract potentials.

\section*{Acknowledgment}
 S.G. was supported by the Special Postdoctoral Researchers Program of RIKEN and iTHEMS Program.
The authors thank K. Yazaki and T. Abe for fruitful discussions on the $\Delta\Delta$ system and its
transition potential to $NN$ system.	T.H., S.A. and T.D. was partially supported by JSPS Grant No. JP18H05236.
S.A. was partially supported by JSPS Grant No. JP16H03978. T.D. was partially supported by JSPS Grant No. JP19K03879 and JP18H05407.
The lattice QCD calculations have been performed on HOKUSAI supercomputers at RIKEN. This work was partially supported 
by “Priority Issue on Post-K computer” (Elucidation of the Fundamental Laws and Evolution of the Universe),
“Program for Promoting Researches on the Supercomputer Fugaku” (Simulation for basic
science: from fundamental laws of particles to creation of nuclei) and Joint Institute for
Computational Fundamental Science (JICFuS).

\appendix
\section{Transition from $\Delta\Delta$ to $NN$}
\label{Sec:TransitionPot}
The threshold of the $NN$ system ($J=3$) in higher partial waves, $^3D_3$ and $^3G_3$, are below the quasi-bound state of $\Delta\Delta$ system.
 In the main text, we  have neglected such transition and derived the single-channel $\Delta\Delta$ potential in the $S$-wave.
To estimate the magnitude of the the decay rate from the quasi-bound state to the $NN$ scattering states, 
 let us calculate the transition potential $V^{NN;\Delta\Delta}(\vec{r})$ 
 by using the general operator form in $I=0$ \cite{Wiringa:1984tg, okubo1958velocity},
 \al{
 V^{NN;NN}(\vec{r}) =& V^{NN;NN}_0(r) + V^{NN;NN}_{\sigma} (r)\vec{\sigma}_1\cdot \vec{\sigma}_2\notag   \\
 &+ V^{NN;NN}_T(r)S_{12}^{\sigma}  \notag \\
 V^{NN;\Delta\Delta}(\vec{r})  =&  V_{S}^{NN;\Delta\Delta}(r) \vec{S}_1 \cdot \vec{S}_2
+V^{NN;\Delta\Delta}_T(r)S_{12}^{S},
}
where $\vec{S}_i (i=1,2)$ is the transition operator from the spin-3/2 state to the spin-1/2 state\footnote{The definition of $\vec{S}$ corresponds to that of $\vec{S}^\dagger$ in Ref.\cite{Wiringa:1984tg}}, and $S_{12}^{A}~(A=\sigma, S)$
is the tensor operator associated with  $\vec{\sigma}$ and $\vec{S}$, respectively:
\al{
S_{12}^{A} &\equiv 3\frac{\left(\vec{A}_1\cdot \vec{r}\right)\left(\vec{A}_2\cdot \vec{r}\right)}{r^2}-\vec{A}_1\cdot \vec{A}_2.}
 For $NN$ system with $s=1$ and $I=0$,  we have $\vec{\sigma}_1\cdot \vec{\sigma}_2 =1$, so that
  $V^{NN;NN}_0(r)$ and $V^{NN;NN}_\sigma(r) \vec{\sigma}_1\cdot \vec{\sigma}_2$ 
  are combined  into
 \al{
V^{NN;NN}_{C}(r) \equiv V^{NN;NN}_0(r) + V^{NN;NN}_{\sigma} (r).}

The potentials, $V^{NN;NN}(r)$ and $V^{NN;\Delta\Delta}(r)$, appear
  in  the coupled channel equations between $NN$ and $\Delta\Delta$ \cite{Aoki:2012bb} 
\al{
&\left(\frac{\nabla^2}{m_N}-\frac{\partial}{\partial t} + \frac{1}{4m_N}\frac{\partial ^2}{\partial t^2}\right)R_J^{NN}(\vec{r},t) \notag \\
&= V^{NN;\Delta\Delta}(\vec{r})R_J^{\Delta\Delta}(\vec{r},t)+V^{NN;NN}(\vec{r})R_J^{NN}(\vec{r},t),
\label{eq:coupled_eq}}
where
$R_J^{NN}(\vec{r},t)$ is given by
\al{	
& R_J^{NN}(\vec{r},t) \notag \\
&= \left<0\right| \left[ N N\right]^{(s=1, I=0)}_J(\vec{r},t) \bar{J}_{\Delta\Delta}^{(s', I=0)}(0)\left|0\right>/e^{-2m_N t} ,
}
with $\left[ N N\right]^{(s=1, I=0)}_J(\vec{r},t)$ being the $NN$ operator with $s=1$, $I=0$, and $J=1,3,$ and $\bar{J}_{\Delta\Delta}^{(s', I=0)}(0)$ being the $\Delta\Delta$ source operator constructed from wall-type quark source 
with internal spin $s' = J$. $R_J^{\Delta\Delta}(\vec{r},t)$ is defined to include the wave function renormalization factor ($Z$-factor) and the kinetic correction factor to compensate the threshold energy difference between $\Delta\Delta$ and $NN$ \cite{Sasaki:2015ifa,Sasaki:2019qnh}.

To extract the potentials from Eq. (\ref{eq:coupled_eq}), we have to utilize $R_J^{NN}(\vec{r},t)$ and $R_J^{\Delta\Delta}(\vec{r},t)$ with given $J$.
Since our $\Delta\Delta$ source operator with internal spin $s'$ is invariant under the $A_1^+$ projection, it contains not only $l=0$ but also $l\ge 4$. Therefore,  it couples to the multiple total angular momenta, $J=s', |s'-4|,|s'-4|+1, \dots $. 
To construct the $NN$-$\Delta\Delta$ correlation with given $J$, we  employ the Misner's projection, where each $(l, l_z)$ contribution can be 
 obtained separately by using points inside the shell that are not connected with each other under the cubic transformation \cite{Misner:1999ab, Miyamoto:2019jjc}. 
For the sink operator with the internal spin $s$, we perform the  $(l, l_z)$ projection by Misner's method and 
have constructed  $J$-projection using  appropriate Clebsch-Gordan coefficients.
 
In principle, we can determine  the four potentials, $V^{NN;NN}_{C}(r)$,   $V^{NN;NN}_T(r)$, $V_{S}^{NN;\Delta\Delta}(r)$, and $V^{NN;\Delta\Delta}_T(r)$,  from the four independent equations obtained by the projection of (\ref{eq:coupled_eq}) into $l=0$ ($S$-wave) and $l=2$ ($D$-wave) components in $J=1$ and $l=2$ ($D$-wave) and $l=4$ ($G$-wave) components in $J=3$.
In practice, however, due to large statistical fluctuations of the $l=4$ component, we cannot determine them precisely.

Alternatively, by assuming that the spin-spin part of the transition potential, $V_{S}^{NN;\Delta\Delta}(r)\vec{S}_1 \cdot \vec{S}_2$, which cannot make the transition from $S$-wave
to higher partial waves, is negligibly small, we have extracted the remnant three potentials 
from the $l=0$ and $l=2$ components in $J=1$ and the $l=2$ component in $J=3$. Again, we have used Misner's projection.

\begin{figure}[tb]
\begin{center}
\includegraphics[keepaspectratio, scale=0.2]{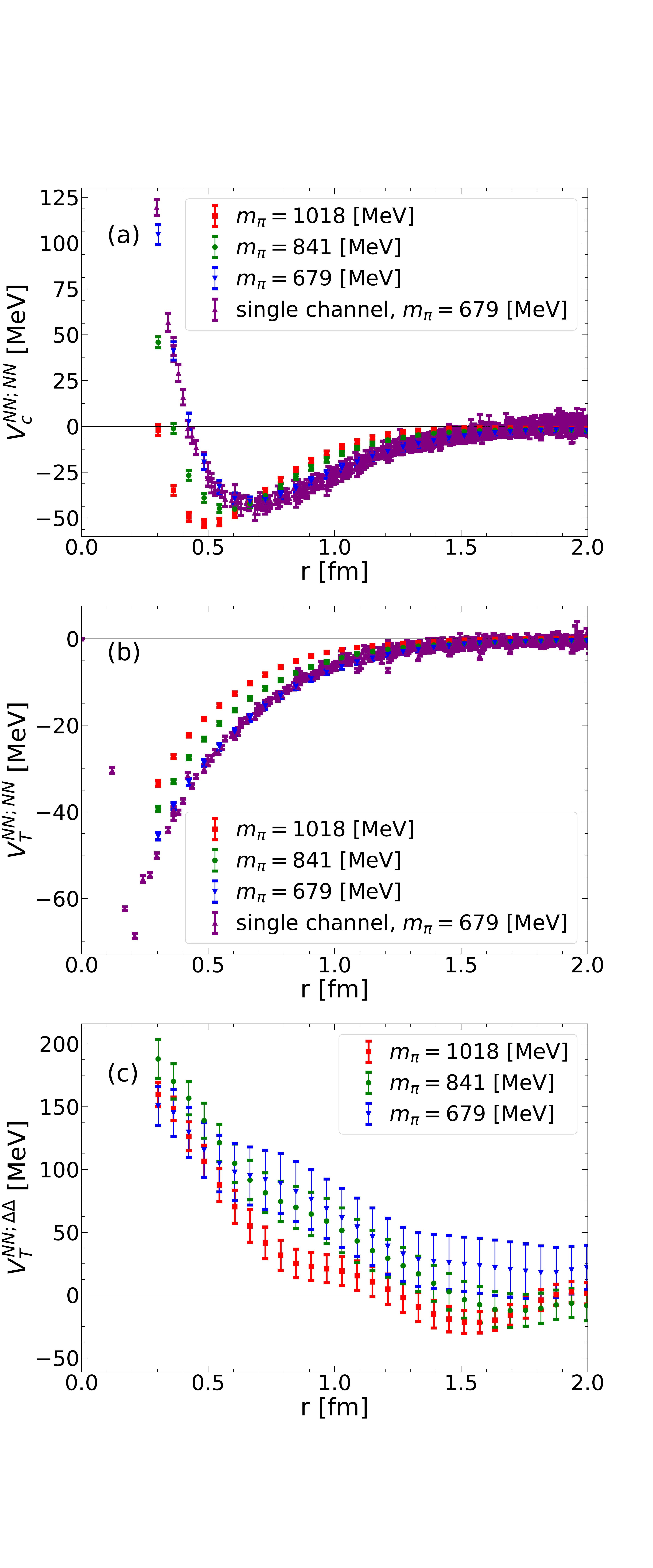} \\
\caption{The central part $V^{NN;NN}_{C}(r)$ and the tensor part $ V^{NN;NN}_T(r)$ of the diagonal potential in $NN$ system 
and the tensor part of the transition potential $V^{NN;\Delta\Delta}_T(r)$ from $\Delta\Delta$ to  $NN$ at $\kappa_{uds} = 0.13710, 0.13760, 0.13800$ corresponding to $m_{\pi}=1018 \mathrm{MeV},  841 \mathrm{MeV}, 679 \mathrm{MeV},$ and $t/a = 10$. The single channel results  for the central potential and the tensor potential at $\kappa_{uds} = 0.13800$ obtained by using $NN$ source in the conventional method are taken from Ref. \cite{Inoue:2011ai}.}
\label{fig:V_NN}
\end{center}
\end{figure}

Shown in Fig.\ref{fig:V_NN}(a)-(c) are the quark mass dependence of the three potentials, $V^{NN;NN}_{C}(r),   V^{NN;NN}_T(r), 
V^{NN;\Delta\Delta}_T(r),$ at $t/a=10$. In Fig. \ref{fig:V_NN}(a) -(b), we observe that the central potential $V^{NN;NN}_{C}(r)$ 
and the tensor potential $V^{NN;NN}_T(r)$
show the qualitatively similar behavior of the phenomenologically well-known potential in the spin-triplet channel of $NN$ system: the short-range repulsion and the intermediate-range
and long-range attraction for the central potential and the all-range negative tensor potential. 
Furthermore, we find that all the results obtained by the coupled channel equations using $\Delta\Delta$ sources in $J=1$ and $J= 3$ are nearly identical with the previous results obtained by the single-channel equation using $NN$ source in $J=1$ \cite{Inoue:2010es,Inoue:2011ai}.  
In Fig. \ref{fig:V_NN}(a) -(b), we also show the results from the single-channel equation at $\kappa_{uds}=0.13800$ corresponding to $m_\pi = 679 \mathrm{MeV }$, taken from Ref. \cite{Inoue:2011ai} (where $m_\pi = 672 \mathrm{MeV}$ is quoted due to the different statistics and fit-range). This good agreement implies that the analysis of the three potentials by neglecting the spin-spin part of the transition potential works well
\footnote{Having neglected the tensor part instead of the spin-spin part, the obtained central potential and tensor potential in $NN$ system are completely different from the previous results in Ref. \cite{Inoue:2011ai}. Even the short-range repulsion cannot be found.  }.

In Fig.\ref{fig:V_NN}(c), we find that the tensor part of the transition potential  increases significantly as $r$ decreases for all the quark masses,
while it has relatively large statistical errors compared with the other potentials. 
Using the transition potential, we then have estimated the decay rate at $J=3$ 
from the quasi-bound state of the
$\Delta\Delta$ system in the $S$-wave to $NN$ in the $D$-wave given by

\al{
\Gamma 
 \simeq&\int \frac{d^3k_1}{\left(2\pi\right)^3}\int \frac{d^3k_2}{\left(2\pi\right)^3}(2\pi)^4\delta^4(k_1^\mu+k_2^\mu - K^\mu) \notag \\
&\times \frac{6}{5}\left|\int r^2dr \bar{\psi}_{{}^3D_3}^{NN} (r)V_T^{NN;\Delta\Delta}(r)\bar{\psi}_{{}^7S_3}^{\Delta\Delta}(r)\right|^2}
with $K^\mu \simeq (2m_\Delta - B_{\Delta\Delta}, {\bm 0})$ and $\bar{\psi}_{{}^3D_3}^{NN} (r)$ and $\bar{\psi}_{{}^7S_3}^{\Delta\Delta}(r)$ being radial wave function of $NN$ scattering state in ${}^3D_3$ channel 
and that of the $\Delta\Delta$ quasi-bound state in ${}^7S_3$ channel, respectively.
Here, we have used the transition potential at $t/a= 10$ by fitting the two $r$-Gaussian form,
$V_T^{NN;\Delta\Delta} (r) = \sum_{i=1}^2 p_i r\exp\left[-\left(r/q_i\right)^2\right]$ with fitting parameters $p_i, q_i~ (i=1,2)$, 
and the $\Delta\Delta$ wave function by solving the Schr\"odinger equation using the central potential at $t/a=10$.
For the sake of simplicity, we have employed the free radial wave function for the $NN$ scattering state,
$\bar{\psi}^{NN}_{{}^3D_3}(r) = -\sqrt{10\pi}j_2(kr),$ with $j_2(kr)$ being the spherical Bessel function of order two. 
This results in $\Gamma = (1.6(6) \mathrm{MeV},5.6(1.7) \mathrm{MeV} ,6.4(1.8) \mathrm{MeV})$ for $m_\pi=(1018 \mathrm{MeV},  841 \mathrm{MeV}, 679 \mathrm{MeV})$. Due to the repulsive interaction, the wave function for the $NN$ scattering state in higher partial waves becomes smaller at short distances, only where the transition potential becomes non-negligible. Therefore, the decay rate is further reduced if more realistic wave function is employed.

\section*{References}
\bibliography{HALQCD}

\end{document}